\documentclass[aps,prb,twocolumn,showpacs,preprintnumbers]{revtex4}
\usepackage{graphicx} 

\begin{document}

\title{Magnetic, Thermal, and Transport Properties of Layered Arsenides
BaRu$_{2}$As$_{2}$ and SrRu$_{2}$As$_{2}$}

\author{R. Nath}
\author{Yogesh Singh}
\author{D. C. Johnston}
\affiliation{Ames Laboratory and
Department of Physics and Astronomy, Iowa State University, Ames, Iowa
50011, USA}

\date{\today}

\begin{abstract}
The magnetic, thermal and transport properties of
polycrystalline BaRu$_{2}$As$_{2}$ and SrRu$_{2}$As$_{2}$ samples
with the ThCr$_{2}$Si$_{2}$ structure were investigated by means of
magnetic susceptibility $\chi(T)$,
electrical resistivity $\rho(T)$, and heat capacity $C_{\rm p}(T)$
measurements. The temperature ($T$) dependence of $\rho$ indicates
metallic character for both compounds with residual resistivity
ratios $\rho$(310~K)/$\rho$(2~K) of $17$ and $5$ for the Ba and Sr
compounds, respectively. The $C_{\rm p}(T)$ results reveal a low-$T$ Sommerfeld coefficient $\gamma = 4.9(1)$ and $4.1(1)$~mJ/mol~K$^{2}$ and Debye temperature $\Theta_{D} = 271(7)$~K and $271(4)$~K for the Ba and
Sr compounds, respectively. The $\chi (T)$ was found to be diamagnetic
with a small absolute value for both compounds. No transitions were found for BaRu$_{2}$As$_{2}$ above 1.8~K\@.  The $\chi(T)$ data for SrRu$_{2}$As$_{2}$ exhibit a cusp at $\sim 200$~K, possibly an indication of a structural and/or magnetic transition.  We discuss the properties of BaRu$_{2}$As$_{2}$ and SrRu$_{2}$As$_{2}$ in the context of other ThCr$_{2}$Si$_{2}$-type and ZrCuSiAs-type transition metal pnictides.
\end{abstract}


\maketitle

\section{Introduction}
The recent discovery of superconductivity in the layered oxypnictide
compound LaFeAsO$_{1-x}$F$_{x}$ with superconducting transition
temperature $T_{c} = 26$ K has generated great
excitement.\cite{kamihara2008} Subsequently a series of compounds
\emph{Ln}FeAsO$_{1-x}$F$_{x}$ (e.g., \emph{Ln} = Ce, Nd, and Sm)
(abbreviated as 1111) have been reported with $T_c$ ranging from
$10$ K to $55$ K,\cite{ren2008, xhchen2008,
gfchen2008, ren2008a} where the high $T_c$ of $55$ K was reached for
SmFeAsO$_{1-x}$F$_{x}$.\cite{ren2008} All these compounds
crystallize in a tetragonal unit cell of ZrCuSiAs structure
type.\cite{quebe2000} Later on another series of compounds
$A_{1-x}$K$_x$Fe$_2$As$_2$ ($A$ = Ba, Sr, Ca, and Eu) (122) (Refs.
\onlinecite{rotter2008a, gfchen2008a, jeevan2008a, sasmal2008,
wu2008}) with the tetragonal ThCr$_2$Si$_2$ structure type\cite{pfisterer1980} was
discovered where the maximum $T_{c}$ achieved was $38$ K.

A common feature of both 1111 and 122 compounds is the identical
FeAs layers separated by the \emph{Ln}O or $A$ layers perpendicular to the
crystallographic c-axis. Undoped metallic parent compounds of both types show
a spin-density wave (SDW) which coexists with a distorted structure at
temperatures $T\lesssim 200$ K.\cite{gfchen2008,
dong2008, klauss2008, rotter2008b, krellner2008, ni2008, yan2008,
ren2008d, jeevan2008b} Superconductivity in both series is sometimes assumed
to be intimately connected with the SDW anomaly in the FeAs
layers.\cite{rotter2008a, dong2008} Electron or hole doping
suppresses both the SDW and structural transition and facilitates
the superconductivity. However, it is still unresolved whether the
structural transition and/or the magnetism associated with the SDW
play a vital role for the occurrence of superconductivity. There
exist a few 122 compounds BaNi$_2$P$_2$,\cite{mine2008}
BaNi$_2$As$_2$,\cite{ronning2008} LaRu$_2$P$_2$,\cite{jeitschko1987}
CsFe$_2$As$_2$, and KFe$_2$As$_2$,\cite{sasmal2008} where even the
undoped compound itself shows superconductivity at low temperatures.
It is of interest to look for further new systems with different
transition metal ions where one can try to achieve an enhanced
$T_{c}$.

\begin{figure}
\includegraphics [width=2.5in]{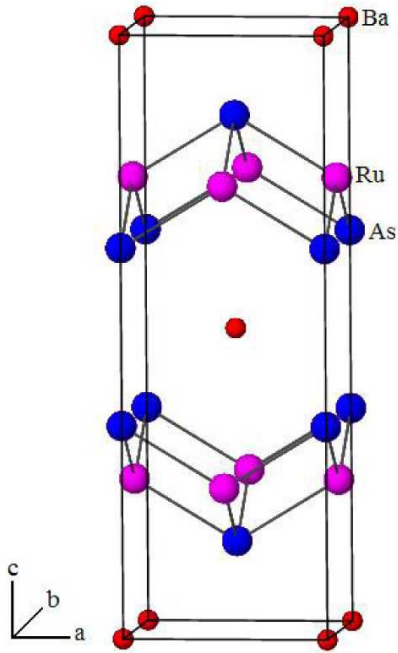}
\caption{\label{structure} (Color online) Crystal structure of
BaRu$_{2}$As$_{2}$ with the tetragonal ThCr$_2$Si$_2$-type structure
showing RuAs layers and Ba layers alternating along the $c$-axis.}
\end{figure}

Motivated by the above progress, we turned our attention towards
Ru-based layered compounds. So far in the Ru series, LaRu$_2$P$_2$
which has the ThCr$_2$Si$_2$-type structure is known to have $T_c=4.1$
K\@.\cite{jeitschko1987} BaRu$_{2}$As$_{2}$ and
SrRu$_{2}$As$_{2}$ compounds also crystallize in
the body-centered tetragonal ThCr$_2$Si$_2$ structure with space
group $I4/mmm$. The reported lattice constants are ($a=4.152$ \AA,
$c=12.238 $ \AA) and ($a=4.168$ \AA, $c=11.179 $ \AA) for the Ba and
Sr compounds, respectively.\cite{jeitschko1987, wenski1986} As shown
in Fig. \ref{structure}, Ru atoms in a square lattice are
coordinated by As to form infinite RuAs layers and the layers are
separated by Ba layers, similar to the $A$Fe$_{2}$As$_{2}$ compounds.
Because BaRu$_{2}$As$_{2}$ and
SrRu$_{2}$As$_{2}$ are isoelectronic to the above undoped
$A$Fe$_{2}$As$_{2}$ compounds, a comparison of the properties of
these two series of compounds is of great interest. A detailed
investigation of the physical
properties of the Ru compounds has not been reported yet. Herein we
report a detailed characterization
of polycrystalline BaRu$_{2}$As$_{2}$ and
SrRu$_{2}$As$_{2}$ by means of magnetic susceptibility $\chi(T)$,
electrical resistivity $\rho(T)$, and heat capacity $C_{\rm p}(T)$
measurements.  We will discuss the properties of BaRu$_{2}$As$_{2}$ and SrRu$_{2}$As$_{2}$ in the context of other ThCr$_{2}$Si$_{2}$-type and ZrCuSiAs-type transition metal pnictides.  From this comparison, it appears that a large Stoner enhancement of the conduction electron spin susceptibility is needed for high $T_{\rm c}$ in this class of materials.

\section{Experimental details}
Polycrystalline samples of BaRu$_{2}$As$_{2}$ and SrRu$_{2}$As$_{2}$
were prepared by solid state reaction techniques using elemental Ba
($99.999$\% pure), Sr ($99.99$\% pure), Ru ($99.9999$\% pure), and
As ($99.999$\% pure). The stoichiometric mixtures in an Al$_2$O$_3$
crucible were sealed inside an evacuated quartz tube. At first, the
elements were heated slowly up to $610$ $^{\circ }$C at a rate of
$80$ $^{\circ }$C/hr, kept there for $10$ hours and then heated up to
$850$ $^{\circ }$C and kept there for $20$ hours. The samples were
then progressively fired at $950$ $^{\circ }$C and $1000$ $^{\circ
}$C for 20 hours, each followed by one intermediate grinding and
pelletization. For the final firing at $1000$ $^{\circ }$C, the
pellets were wrapped in a Ta foil before sealing in the quartz tube.
All the sample handling was carried out inside a He-filled glove
box.    Our repeated attempts
to grow crystals using Sn, Pb, and In fluxes followed by slow
cooling failed.

\begin{figure}
\includegraphics [width=3in]{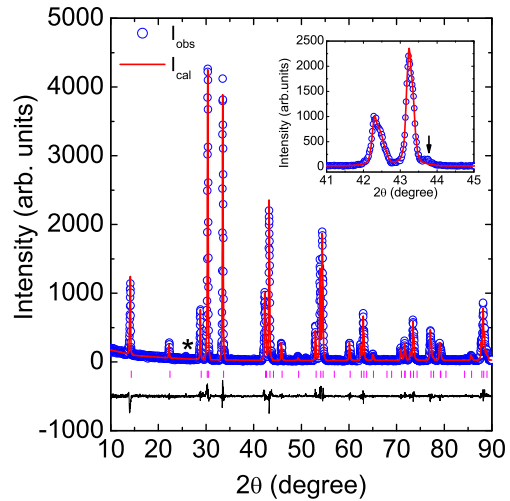}
\caption{\label{xrd}
(Color online) X-ray powder diffraction pattern (open circles)
at room temperature for BaRu$_{2}$As$_{2}$. The solid line
represents the Rietveld refinement fit with the ThCr$_2$Si$_2$
structure and $I4/mmm$ space group. The impurity peak corresponding to BaCO$_{3}$ is marked by a star. Inset: small section of the
x-ray powder pattern is magnified to highlight the unidentified
impurity peak marked by an arrow, which might be due to unreacted Ru metal.}
\end{figure}

The samples were characterized using a Rigaku Geigerflex powder
X-ray diffractometer and Cu K$_{\alpha}$ radiation
($\lambda_{av}=1.54182$ \AA ). The
powder pattern evidenced almost single phase material with a
small impurity peak of about $2$ and $3$ percent relative intensity
for the Ba and Sr compounds,
respectively, which is suspected to arise from unreacted Ru.
Unlike the Sr compound, for the Ba compound there appears another tiny impurity peak at the position expected for the strongest peak
of BaCO$_{3}$ at $2\theta \approx 25^{\circ}$
(marked by a star) which has about $1.5\%$ relative intensity. These impurities should not measurably affect the $\chi(T)$ or $C_{\rm p}(T)$ data but may have an unknown effect on the $\rho(T)$ data.  Rietveld
refinements of the data were carried out using the GSAS
package.\cite{larson2000} Figure~\ref{xrd} shows the Rietveld
refinement fit to the x-ray powder diffraction pattern for
BaRu$_{2}$As$_{2}$. All the peaks except for
the above impurity peak (see inset of Fig.~\ref{xrd}) could be
indexed and fitted based on the ThCr$_2$Si$_2$ structure with
$I4/mmm$ space group. The goodness of the fit is $5.3\%$ and
$7.5\%$ for the Ba and Sr compounds, respectively. The obtained
lattice parameters are [$a=4.15248(8)$~\AA, $c=12.2504(3)$~\AA]
and [$a=4.1713(1)$~\AA, $c=11.1845(4)$~\AA] for the Ba and Sr
compounds, respectively. These values are close to the above
previously reported ones.\cite{jeitschko1987}  Our refined $z$ parameters for the As atoms are 0.3527(1) for BaRu$_2$As$_2$ and 0.3612(2) for SrRu$_2$As$_2$.

The magnetization $M$ and magnetic susceptibility $\chi (T) \equiv M/H$, where $H$ is the applied magnetic field, were measured in
the temperature $T$ range \mbox{$1.8$ K $\leq $ $T$ $\leq $ $300$ K} on polycrystalline samples in a commercial
Quantum Design SQUID (superconducting quantum interference device)
magnetometer. DC resistivity $\rho(T)$ was measured
using a standard 4-probe technique by applying a current of $5$ mA,
and heat capacity $C_{\rm p}(T)$ was
measured on a small piece of sample with mass about $8$ mg. Both
$\rho(T)$ and $C_{\rm p}(T)$ measurements were performed on sintered
pellets using a Quantum Design Physical Property Measurement System.

\section{Experimental Results and Analysis}

\begin{figure}
\includegraphics [width=3in]{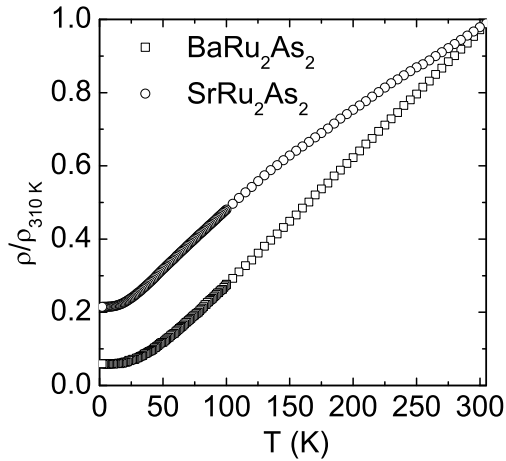}
\caption{\label{resistivity}
Normalized DC electrical resistivity $\rho$ versus
temperature $T$ of BaRu$_{2}$As$_{2}$ and SrRu$_{2}$As$_{2}$. The
room temperature resistivity $\rho(310$ K) values are
170~$\mu \Omega$~cm and 10.7 m$\Omega$~cm for BaRu$_{2}$As$_{2}$
and SrRu$_{2}$As$_{2}$, respectively.  The unexpectedly large value for SrRu$_{2}$As$_{2}$ may arise from a high porosity of the sample.}
\end{figure}

Figure~\ref{resistivity} shows the temperature dependence of
resistivity $\rho(T)$ for BaRu$_{2}$As$_{2}$ and SrRu$_{2}$As$_{2}$. With decreasing
temperature, $\rho(T)$ decreases for both compounds to a
residual resistivity at $2$~K of about $9.7$ $\mu \Omega$~cm and
$2300$~$\mu \Omega$~cm for the Ba and Sr compounds, respectively. This
type of temperature dependence suggests metallic behavior of the
compounds. We did not observe any clear anomaly that might be associated
with an SDW down to $2$ K\@. The residual resistivity
ratio \mbox{$\rho(310$ K)/$\rho(2$ K)} was found to be about $17$ and $5$
for the Ba and Sr compounds, respectively. These values are comparable to
those reported for polycrystalline (Ba,Sr)Fe$_2$As$_2$.\cite{rotter2008b,
krellner2008}

\begin{figure}
\includegraphics [width=3in]{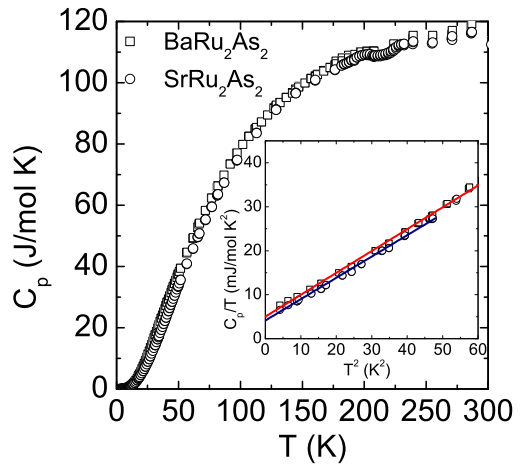}
\caption{\label{cp} (Color online) Heat capacity $C_{\rm p}$ vs.
temperature $T$ for BaRu$_{2}$As$_{2}$ and SrRu$_{2}$As$_{2}$.
The inset shows
$C_{\rm p}/T$ vs. $T^{2}$ at low temperatures and the two solid
lines are the respective linear fits.}
\end{figure}

The heat capacity $C_{\rm p}$ vs. $T$ at zero field for BaRu$_{2}$As$_{2}$ and SrRu$_{2}$As$_{2}$ is shown in
Fig.~\ref{cp}. We did not observe any clear anomaly associated with a SDW or
structural transition down to $2$ K\@. However, the small anomaly at about 200~K for SrRu$_{2}$As$_{2}$ may be a real effect in view of the cusp at the same temperature found in the measurement of $\chi(T)$ below.  The value of $C_{\rm p}$
at room temperature is about $120$ J/mol K which is close to the
Dulong Petit lattice heat capacity value
$C_{\rm p}=15R \simeq 125$~J/mol K expected for our
compounds,\cite{kittel1966} where $R$ is the molar gas constant. As
shown in the inset of Fig.~\ref{cp}, $C_{\rm p}(T)/T$ vs.~$T^{2}$ is
almost linear at low-$T$ ($T<8$ K) and was fitted
by the expression $C_{\rm p}/T = \gamma+\beta T^{2}$, where $\gamma$
is the Sommerfeld coefficient of electronic heat capacity and the
second term accounts for the lattice contribution with coefficient
$\beta$. The resultant $\gamma$ and $\beta$ values are [$4.9(1)$
mJ/mol K$^{2}$ and $0.49(4)$ mJ/mol K$^{4}$] and [$4.1(1)$
mJ/mol K$^{2}$ and $0.49(2)$ mJ/mol K$^{4}$] for the Ba and Sr compounds,
respectively.

The density of states at the Fermi energy for both spin directions
$N(E_{F})$ can be estimated using the value of $\gamma$ in the
following relation,\cite{kittel1966}

\begin{equation}
\gamma =\frac{\pi ^{2}}{3}%
k_{\rm B}^{2}N(E_{F})\left( 1+\lambda _{\rm ep}\right)  \label{gama}
\end{equation}%
where $\lambda _{\rm ep}$ is the electron-phonon coupling constant. As a
first approximation we set $\lambda _{\rm ep}=0$, which gives
$N(E_{F}) = 2.1(1)$~states/(eV~f.u.) (f.u. stands for formula
unit) and $1.7(1)$~states/(eV~f.u.) for the Ba and Sr
compounds, respectively. These densities of states are comparable
with our previously reported value of $2.0(4)$~states/(eV~f.u.) for
BaRh$_{2}$As$_{2}$ estimated in the same way from heat capacity
measurements.
\cite{singh2008} In BaRh$_{2}$As$_{2}$, band structure calculations
indicate that the maximum contribution to $N(E_{F})$ comes from
the Rh $4d$ states.\cite{singh2008} From $N(E_{F})$ one can calculate
the bare Pauli paramagnetic spin susceptibility of the conduction
carriers $\chi_{\rm P}$ using\cite{kittel1966}
\begin{equation}
\chi_{\rm P}=\mu_{\rm B}^{2} N(E_{F})
\end{equation}%
where $\mu_{\rm B}$ is the Bohr magneton. This gives
$\chi_{\rm P}\simeq
6.8 \times 10^{-5}$ cm$^3$/mol and $5.5 \times 10^{-5}$ cm$^3$/mol
for the Ba and Sr compounds, respectively. These values are comparable
to that found in BaRh$_2$As$_2$.\cite{singh2008} From the value
of $\beta$ one
can also estimate the Debye temperature $\Theta _{D}$ using the
expression,\cite{kittel1966}
\begin{equation}
\Theta _{D}=\left( \frac{%
12\pi ^{4}Rn}{5\beta }\right) ^{1/3}  \label{theta}
\end{equation}%
where $n$ is the number of atoms
per formula unit ($n=5$ for our compounds). The above $\beta$ values
yield $\Theta _{D} = 271(7)$~K and $271(4)$~K for the Ba and Sr
compounds, respectively, which are comparable to the values of
$\sim 280$~K (Ref. \onlinecite{an2009}) and $246(3)$~K
(Ref.~\onlinecite{singh2009}) reported
for isostructural BaMn$_2$As$_2$ but are larger than the value of
$171(2)$~K reported for BaRh$_2$As$_2$.\cite{singh2008}

\begin{figure}[t]
\includegraphics [width=3in]{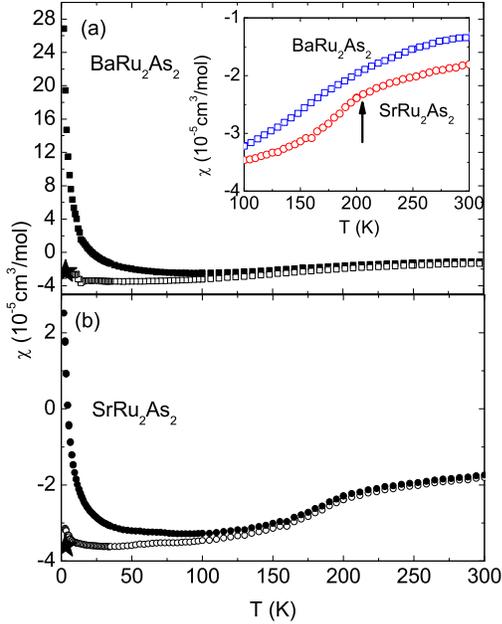}
\caption{\label{susc} (Color online) Magnetic susceptibility 
$\chi \equiv M/H$
of (a) BaRu$_{2}$As$_{2}$ and (b) SrRu$_{2}$As$_{2}$ versus
temperature $T$
(solid symbols) measured in an applied magnetic field $H=1$~T\@.  The magnetization $M$ data are corrected for the contributions from ferromagnetic impurities as determined from the $M(H)$ isotherm data in Fig.~\ref{MH} (see text).  The intrinsic $\chi (T)$ after correction for the additional contribution of paramagnetic impurities (see text) is also shown (open symbols). The star symbols are the $\chi_{0}$ values in Eq.~(\ref{EqM(H)}) estimated
from the analysis of magnetization versus field isotherms at low temperatures. In the inset of (a) the intrinsic $\chi (T)$ data are magnified for both compounds and the arrow points to the anomaly around $200$ K for SrRu$_{2}$As$_{2}$.}
\end{figure}

For both the magnetic susceptibility $\chi(T)$ and magnetization $M(H,T)$ measurements we carried out, the data were corrected for the contribution of the empty sample holder, which was not negligible.  The $\chi \equiv M/H$ as a function of
temperature in a field $H=1$~T is
shown in Fig.~\ref{susc}, where the $M$ data are corrected for the contributions from ferromagnetic impurities as determined from the $M(H)$ isotherm data in Fig.~\ref{MH}  below. $\chi (T)$ at high temperatures is
negative and weakly temperature dependent. At low temperatures,
$\chi (T)$ shows a Curie-like upturn. For the Ba compound this
upturn is much stronger than
the Sr one. This upturn is attributed to extrinsic paramagnetic
impurities and/or magnetic defects in the samples, as now discussed.

\begin{figure}
\includegraphics [width=3.5in]{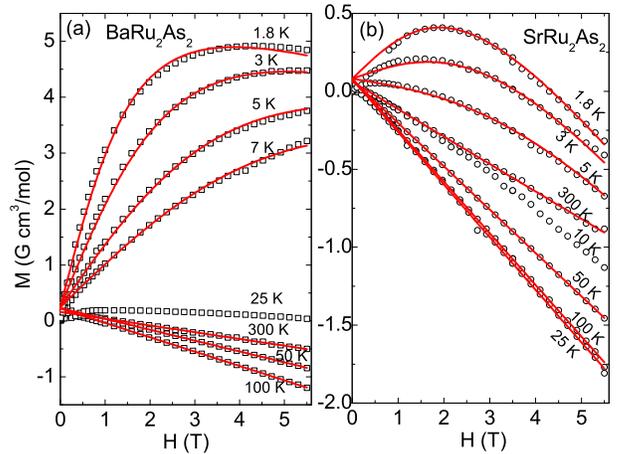}
\caption{\label{MH} (Color online) Magnetization $M$ versus
field $H$ isotherms for
(a) BaRu$_{2}$As$_{2}$ and (b) SrRu$_{2}$As$_{2}$ at different
temperatures. For SrRu$_{2}$As$_{2}$ at $1.8$~K, we were not
able to collect
data below $5$~kOe where the sum of the signals from the sample and sample holder is nearly zero. 
Solid curves are the fits by Eq.~(4) at $1.8$~K, $3$~K, $5$~K
and $7$~K for the Ba compound and at $1.8$~K, $3$~K and $5$~K for the Sr compound.  The straight lines are fits to the data for $H \ge 1$~T at $\geq 50$~K for the Ba compound and at $\geq 25$~K for the Sr compound by $M = M_{\rm s} + \chi H$.}
\end{figure}

Magnetization $M$ as a function of applied magnetic field $H$
was measured at
different temperatures. Figure~\ref{MH} shows the $M(H)$ isotherms
at different temperatures for both compounds measured up to
$H = 5.5$~T\@.  For both compounds $M(H)$ shows a negative curvature below a field of 1~T at all temperatures due to the saturation of ferromagnetic impurities. To
estimate the saturation magnetization $M_{s}$ of the ferromagnetic
impurities, we fitted magnetization isotherms at high temperatures
[$\geq 50$~K for BaRu$_2$As$_2$ and $\geq 25$~K for SrRu$_2$As$_2$]
to a straight line ($M_{\rm s}+\chi H$) above 1~T, as shown by the
straight line fits in Fig.~\ref{MH}. The $M_{\rm s}$ was found to be
nearly independent of $T$ with values of about $0.247$~G~cm$^3$/mol and
$0.081$~G~cm$^3$/mol at low temperatures for the Ba and Sr compounds, respectively.  The near constant values of $M_{\rm s}$ versus temperature indicate that the Curie temperature(s) of the ferromagnetic impurities are significantly above 300~K\@.  The low-$T$ $M_{\rm s}$ values correspond to the contributions of the equivalent of $20$ molar ppm and $6.6$ molar ppm of Fe metal impurities to the $M$ of the Ba and Sr compounds, respectively.  The plotted magnetic susceptibilities at $H = 1$~T as given above by the filled symbols in Fig.~\ref{susc} are given by $\chi(T) = [M(T) - M_{\rm s}(T)]/H$.

For a quantitative estimation of the paramagnetic
impurity contribution giving rise to the upturns in the susceptibilities at low temperatures, we fitted our $M(H)$ data for
$1$~T $\leq $ $H$ $\leq 5.5$~T at
$1.8$~K, $3$~K, $5$~K and $7$~K for the Ba compound and $1.8$~K, $3$~K and $5$~K for the Sr compound simultaneously by the equation
\begin{equation}
M=M_{\rm s}+\chi_{0} H+f_{\rm imp}N_{\rm A}g_{\rm imp}\mu _{\rm B}
S_{\rm imp}B_{S_{\rm imp}}(x)
\label{EqM(H)}
\end{equation}
where $M_{\rm s}$ is the above-determined low-temperature ferromagnetic impurity
saturation value, $f_{\rm imp}$ is the molar fraction of the
impurities, $N_{\rm
A}$ is Avogadro's number, $g_{\rm imp}$ is the impurity $g$-factor,
$S_{\rm imp}$ is the impurity spin, $B_{S_{\rm imp}}(x)$ is the
Brillouin function,\cite{kittel1966} and $\chi_{0}$ is the
intrinsic susceptibility
of the sample. The modified argument of the Brillouin function
is $x=g_{\rm
imp}\mu_{\rm B}S_{\rm imp}H/[k_{\rm B}(T-\theta_{\rm imp})]$ where
$\theta_{\rm imp}$ is the Weiss temperature due to impurity
interactions. To reduce the number
of fitting parameters, the impurity $g$-factor was set to $g_{\rm imp}=2$. The
obtained fitting parameters ($\chi _{0}$,
$f_{\rm imp}$, $S_{\rm imp}$, and $\theta_{\rm imp}$) are
[$-2.3(1)\times 10^{-5}$~cm$^3$/mol, $0.0284(3)$~mol\%, $1.85(3)$, and
$-0.46(6)$~K] and [$-3.62(6)\times 10^{-5}$~cm$^3$/mol,
$0.0092(1)$~mol\%, $1.62(4)$, and $-1.2(1)$~K] for the Ba and Sr compounds,
respectively. The Curie constant $C_{\rm imp}$ of the paramagnetic
impurities was calculated
using $C_{\rm imp}=f_{\rm imp}N_{\rm A}g_{\rm imp}^{2}\mu _{\rm
B}^{2}S_{\rm imp}(S_{\rm imp}+1)/3k_{\rm B}$ which yields $C_{\rm
imp} \approx 0.754\times 10^{-3}$~cm$^3$ K/mol and
$0.195\times 10^{-3}$~cm$^3$ K/mol for the Ba and Sr compounds,
respectively.
Our intrinsic $\chi (T)$ data that are corrected for both the
ferromagnetic impurity and paramagnetic impurity contributions are
shown in Fig.~\ref{susc} as open symbols. The low-$T$ $\chi_{0}$
values obtained from the magnetization isotherm analysis are
also plotted as filled stars in Fig.~\ref{susc} and are, of course, in agreement with the intrinsic $\chi(T)$ data.

From the open symbols in Fig.~\ref{susc}, the intrinsic susceptibilities of
BaRu$_2$As$_2$ and SrRu$_2$As$_2$ are diamagnetic over the
whole $T$ range, becoming somewhat more negative with
decreasing $T$. A diamagnetic susceptibility is not unprecedented for a transition metal compound, as seen, e.g., for LaRu$_2$P$_2$ [Ref.~\onlinecite{jeitschko1987}] and OsB$_2$ and RuB$_2$.\cite{Singh2007}  As shown in the inset of Fig.~\ref{susc}, the
intrinsic $\chi (T)$ of SrRu$_2$As$_2$ shows a (reproducible)
small cusp
around $200$ K in contrast to the smooth behavior
observed in BaRu$_2$As$_2$. This cusp for SrRu$_2$As$_2$ is
qualitatively similar to the cusp seen for 1111 and 122 parent
compounds and attributed to structural/SDW transitions.
\cite{gfchen2008, dong2008, klauss2008, rotter2008b, %
krellner2008, ni2008, yan2008, ren2008d, jeevan2008b}
The temperature of the cusp is similar to the temperature of the small anomaly in $C_{\rm p}(T)$ in Fig.~\ref{cp}, so the latter anomaly may not be an artifact.

\section{Discussion}

The intrinsic $\chi (T)$ of a metal can be written as $\chi = \chi_{\rm D}
+ \chi_{\rm VV} + \chi_{\rm P}$, where $\chi_{\rm D}$ includes the orbital 
diamagnetism of the core electrons ($\chi_{\rm core}$) and the orbital
Landau diamagnetism ($\chi_{\rm L}$) of the conduction electrons,
$\chi_{\rm VV}$ is the orbital Van Vleck paramagnetism, and $\chi_{\rm P}$
is the Pauli spin paramagnetism of the conduction electrons.  For an
extended system it is difficult to calculate $\chi_{\rm D}$ due to
intercell currents. Nevertheless one can roughly estimate the
$\chi_{\rm core}$ of the compounds assuming an ionic picture, where
(Ba or Sr), Ru, and As are in $2+$, $2+$, and $3-$ oxidation states,
respectively.\cite{Magnetochemistry} This estimate will give the
upper limit to the $\chi_{\rm D}$. In this way $\chi_{\rm core}$ was
calculated to be $-1.6 \times 10^{-4}$ cm$^{3}$/mol and $-1.43
\times 10^{-4}$ cm$^{3}$/mol for the Ba and Sr compounds,
respectively. Since $\chi_{\rm P}$ is positive,
when added to the negative $\chi_{\rm core}$, the result is a
reduced positive value or even a negative value of $\chi$.
Using the $\chi_{\rm P}$ values obtained from the above heat
capacity data analysis, the sum of $\chi_{\rm core}$ and
$\chi_{\rm P}$ is $\approx -9\times 10^{-5}$
cm$^3$/mol for both
compounds. This value is somewhat more
negative than the intrinsic values in Fig.~\ref{susc},
suggesting that the Van Vleck paramagnetic orbital
susceptibility $\chi_{\rm VV}$ and/or Stoner enhancement of $\chi_{\rm P}$ are not negligible in these
compounds. This value is
also of the same order of magnitude as has been estimated for
BaRh$_{2}$As$_{2}$.\cite{singh2008}

In the following discussion we relate the properties of the (Sr,Ba)Ru$_{2}$As$_{2}$ compounds with those of other ThCr$_{2}$Si$_{2}$-type and ZrCuSiAs-type pnictides and consider their superconducting transition temperature $T_{\rm c}$ or lack thereof.  Lee and coworkers\cite{Lee2008, Lee2008b} and Zhao and coworkers\cite{Zhao2008} found an interesting correlation for a wide range of parent compounds Ba(Fe,Ni)$_2$(P,As)$_2$ and \emph{Ln}(Fe,Ni)(P,As)O  where \emph{Ln} is a rare earth element: the highest $T_{\rm c}$ occurred  for the doped materials in which the respective FeAs$_4$ tetrahedra were least distorted.\cite{Lee2008, Lee2008b, Zhao2008}  Within a \emph{MPn}$_4$ tetrahedron where $M$ is the transition metal and \emph{Pn} = P or As, there is a two-fold \emph{Pn-M-Pn} bond angle where the two \emph{Pn} atoms are on the same side of the $M$ atom layer along the $c$ axis, and there is a four-fold \emph{Pn-M-Pn} bond angle where the two \emph{Pn} atoms are on opposite sides of the $M$ layer (see Fig.~\ref{structure}). ÊThe angle plotted in Refs.~\onlinecite{Lee2008} and~\onlinecite{Lee2008b} is the two-fold \emph{Pn-M-Pn} bond angle. Ê

We have calculated the two-fold As-Ru-As bond angles from our structural data for the (Sr,Ba)Ru$_{2}$As$_{2}$ compounds and also for the As-Rh-As bond angle for BaRh$_{2}$As$_{2}$.\cite{singh2008}  For the body-centered-tetragonal BaFe$_{2}$As$_{2}$-type and the primitive tetragonal LaFeAsO-type structures, the two-fold and four-fold As-Fe-As bond angles are given by
\begin{eqnarray}
\theta_2 &=& {\rm arccos}\left[\frac{-\frac{a^2}{4} + (z - \alpha)^2 c^2}{r^2}\right] \ \ \ \ ({\rm 2-fold})\nonumber\\
\theta_4 &=& {\rm arccos}\left[\frac{-(z - \alpha)^2 c^2}{r^2}\right] \ \ \ \ \ \ \ \ \ \ \ ({\rm 4-fold})\nonumber\\
{\rm where} \label{Eqtheta}\\
r^2 &=& \frac{a^2}{4} + \left(z - \alpha\right)^2 c^2\nonumber
\end{eqnarray}
and $\alpha = 1/4$ for the BaFe$_{2}$As$_{2}$-type structure and \mbox{$\alpha = 1/2$} for the LaFeAsO-type structure.  Here $a$ and $c$ are the lattice parameters, $z$ is the $z$-axis position parameter of the As atom in the respective structure ($z\approx0.35$ in BaFe$_{2}$As$_{2}$ and $z\approx0.65$ in LaFeAsO), and $r$ is the nearest-neighbor Fe-As distance within an Fe-centered FeAs$_4$ tetrahedron (all four Fe-As nearest-neighbor distances are the same in each of the two structures).  The Fe atoms in both structures form a square lattice where the 4-fold nearest-neighbor Fe-Fe distance in both structures is $d_{\rm Fe-Fe} = a/\sqrt{2}$. 

Using Eqs.~(\ref{Eqtheta}), we find $\theta_2 = 117.6^\circ,\ 118.4^\circ$, and $112.2^\circ$ for BaRu$_{2}$As$_{2}$, SrRu$_{2}$As$_{2}$ and BaRh$_{2}$As$_{2}$, respectively. These bond angles for the (Ba,Sr)Ru$_{2}$As$_{2}$ compounds are significantly larger than the above optimum value of $\approx 110^\circ$ for the Fe(P,As)-based materials and therefore the low $T_{\rm c}$'s $< 1.8$~K for the (Ba,Sr)Ru$_{2}$As$_{2}$ compounds are consistent with this overall behavior.  On the other hand, BaRh$_{2}$As$_{2}$ stands out as an exception: it has the same $\theta_2$ as the high temperature superconducting LaFeAsO-based and CeFeAsO-based compounds with $T_{\rm c} = 28$--40~K (Refs.~\onlinecite{Lee2008} and \onlinecite{Zhao2008}) but is not superconducting.  Therefore, at least one additional characteristic of the materials must be controlling $T_{\rm c}$.  As will be discussed in detail elsewhere,\cite{Johnston2009} the bare nonmagnetic band structure density of states at the Fermi energy $N(E_{\rm F})$ is not correlated with $T_{\rm c}$.  For example, the calculated $N(E_{\rm F})$ for nonsuperconducting BaRh$_{2}$As$_{2}$ and that of superconducting LaFeAsO$_{0.9}$F$_{0.1}$ with $T_{\rm c}$ = 27~K are the same.  The values are 1.76 [Ref.~\onlinecite{singh2008}] and 1.28--2.01 states/(eV-$M$ atom) [Refs.~\onlinecite{mazin2008} and \onlinecite{Wang2009}], respectively, for both spin directions.  

On the other hand, $\chi(300$~K) is large for all FeAs-based compounds with high $T_{\rm c}$, suggesting that Stoner enhancement of the susceptibility may be relevant to the superconducting mechanism.\cite{Johnston2009}  Again using the same examples as above, for BaRh$_{2}$As$_{2}$ one finds a powder averaged $\bar{\chi}$(300~K) = $0.18 \times 10^{-4}\ {\rm cm^3}$/(mol Rh),\cite{singh2008} whereas for LaFeAsO$_{0.9}$F$_{0.1}$ one obtains $\bar{\chi}$(300~K) = $3.3 \times 10^{-4}\ {\rm cm^3}$/(mol Fe),\cite{Klingeler2008} which is a factor of 18 larger.  For LaFeAsO$_{0.9}$F$_{0.1}$, the above $N(E_{\rm F})$ range predicts a bare Pauli conduction electron spin susceptibility of 0.41--0.65 $\times 10^{-4}\ {\rm cm^3}$/(mol Fe), suggesting a Stoner enhancement by a factor of 5 to 8.  Howevever, accurate estimates of the enhancement factor require that orbital contributions to $\chi$ must be corrected for.\cite{Johnston2009}

\section{Conclusion}
We have synthesized and investigated the physical properties of
ThCr$_2$Si$_2$-type polycrystalline BaRu$_{2}$As$_{2}$ and
SrRu$_{2}$As$_{2}$ compounds. Both compounds were found to be
metallic in character. Unlike other similar \emph{isoelectronic}
compounds (Ca,Sr,Ba)Fe$_2$As$_2$, BaRu$_{2}$As$_{2}$ shows no signature of
 a spin density wave or a structural transition from $\rho(T)$, $\chi(T)$ or $C_{\rm p}(T)$ measurements down to $1.8$~K\@. However, a clear cusp in $\chi(T)$ and a hint of one in $C_{\rm p}(T)$ was found at $\sim 200$~K for SrRu$_{2}$As$_{2}$, which may be indicative of a structural and/or magnetic transition.

From analysis of our $C_{\rm p}(T)$ data, the density of states at
the Fermi energy $N(E_{\rm F})$ for (Ba,Sr)Ru$_{2}$As$_{2}$ was estimated to be \mbox{$\sim 1$ state/(eV Ru atom)} for both spin directions and is comparable to that per Rh atom in BaRh$_2$As$_2$. The small and negative values of $\chi (T)$ for both BaRu$_{2}$As$_{2}$ and SrRu$_{2}$As$_{2}$ indicate small or negligible Stoner enhancement of the conduction electron spin susceptibility.  A comparison of $N(E_{F})$ with that of BaRh$_2$As$_2$ suggests that the maximum contribution to $N(E_{F})$ comes from the Ru $4d$ states.
It would be interesting to study the effect of doping of SrRu$_{2}$As$_{2}$ in view of the occurrence of  superconductivity in Ba$_{1-x}$K$_{x}$Fe$_{2}$As$_{2}$, 
Sr$_{1-x}$K$_{x}$Fe$_{2}$As$_{2}$ and BaFe$_{2-x}$Ru$_x$As$_{2}$.\cite{sasmal2008, rotter2008a, Paulraj2009}  Finally, a comparison of the properties of (Ba,Sr)Ru$_{2}$As$_{2}$ and BaRh$_{2}$As$_{2}$ with those of the FeAs-based materials indicates that $N(E_{\rm F})$ for these nonsuperconducting Ru and Rh arsenides is about the same as for FeAs-based compounds with high $T_{\rm c}$.\cite{Johnston2009}  A distinguishing feature of the high $T_{\rm c}$ FeAs-based materials is their large $\chi$ values that evidently reflect signficant Stoner enhancement of the conduction electron spin susceptibility\cite{Johnston2009} as has been pointed out before.\cite{djsingh2008, Kohama2008, Haule2009}  Thus it seems possible that the mechanisms for the Stoner enhancement and for the high $T_{\rm c}$ in the FeAs-based materials may be closely related.

\begin{acknowledgments}
Work at the Ames Laboratory was supported by the Department of
Energy-Basic Energy Sciences under Contract No. DE-AC02-07CH11358.
\end{acknowledgments}

\end{document}